\documentclass[twocolumn,aps,showpacs,floatfix,prc]{revtex4}
\usepackage[dvips]{epsfig}
\begin{document}


\title{Isobaric incompressibility of the isospin asymmetric nuclear matter}

\author{D. N. Basu$^1$, P. Roy Chowdhury$^2$, C. Samanta$^{2,3,4}$}

\affiliation{ $^1$ Variable  Energy  Cyclotron  Centre, 1/AF Bidhan Nagar, Kolkata 700 064, India }
\affiliation{ $^2$ Saha Institute of Nuclear Physics, 1/AF Bidhan Nagar, Kolkata 700 064, India }
\affiliation{ $^3$ Physics Department, Virginia Commonwealth University, Richmond, VA 23284-2000, U.S.A.}
\affiliation{ $^4$ Physics Department, University of Richmond, VA 23173, U.S.A.}

\email[E-mail: ]{dnb@veccal.ernet.in}
\email[E-mail: ]{partha.roychowdhury@saha.ac.in} 
\email[E-mail: ]{chhanda.samanta@saha.ac.in}

\date{\today }

\begin{abstract}

      The isospin dependence of the saturation properties of asymmetric nuclear matter, particularly the incompressibility $K_\infty (X) = K_\infty + K_\tau X^2 + O(X^4)$ at saturation density is systematically studied using density dependent M3Y interaction. The $K_\tau$ characterizes the isospin dependence of the incompressibility at saturation density $\rho_0$. The approximate expression $K_{asy} \approx K_{sym}-6L$ is often used for $K_\tau$ where $L$ and $K_{sym}$ represent, respectively, the slope and curvature parameters of the symmetry energy at $\rho_0$. It can be expressed accurately as $K_\tau=K_{sym}-6L-\frac{Q_0}{K_\infty}L$ where $Q_0$ is the third-order derivative parameter of symmetric nuclear matter at $\rho_0$. The results of this addendum to Phys. Rev. {\bf C 80}, 011305(R) (2009) indicate that the $Q_0$ contribution to $K_\tau$ is not insignificant.

\end{abstract}


\maketitle

      The nuclear symmetry energy (NSE) $E_{sym}$ and its density dependence \cite{Li02} are critical for the understanding of heavy-ion reactions \cite{Li98,Da02,Ba05}, structure of rare isotopes \cite{Br00} and many interesting issues in astrophysics \cite{Li08,Su94,La04,St05}. The determination of the NSE has been a long-standing goal of both nuclear physics and astrophysics and both fields have some promising tools for probing it over a wide density range. However, they all have some limitations and by combining carefully the complementary information from both fields, it is possible to obtain some understanding about the NSE. While significant progress has been made in constraining the $E_{sym}$ at subsaturation densities using terrestrial nuclear laboratory data, still very little is known about the $E_{sym}$ at suprasaturation densities. The isospin dependent part $K_\tau$ of the isobaric incompressibility $K$, the slope $L$ of $E_{sym}$, and $E_{sym}(\rho_0)$, the quantities which can be extracted experimentally, provide information about the density dependent behaviour of $E_{sym}$ around the saturation density $\rho_0$. 

      The isobaric incompressibility for infinite nuclear matter can be expanded in the power series of isospin asymmetry $X$ as $K_\infty (X) = K_\infty + K_\tau X^2 + K_4 X^4 + O(X^6)$ where $X$=$\frac{\rho_n - \rho_p} {\rho_n + \rho_p}$ with $\rho_n$, $\rho_p$ and $\rho$=$\rho_n$+$\rho_p$ being the neutron, proton and nucleonic densities respectively. The magnitude of the higher-order $K_4$ parameter is generally quite small compared to $K_\tau$ \cite{Ch09}. The latter essentially characterizes the isospin dependence of the incompressibility at saturation density and can be expressed as $K_\tau=K_{sym}-6L-\frac{Q_0}{K_\infty}L$ where $L$ and $K_{sym}$ represent, respectively, the slope and curvature parameters of the symmetry energy at the nuclear matter saturation density $\rho_0$ while $Q_0$ is the third-order derivative parameter of the symmetric nuclear matter (SNM) at $\rho_0$. The approximate expression $K_{asy} \approx K_{sym}-6L$ is quite often used for $K_\tau$. In this short report, we study the contribution of $Q_0$ to $K_\tau$.

      The nuclear matter EoS is calculated \cite{BCS08} using the isoscalar and the isovector components of M3Y interaction along with density dependence. The density dependence of the effective interaction, DDM3Y, is completely determined from nuclear matter calculations. The equilibrium density of the nuclear matter is determined by minimizing the energy per nucleon. The energy variation of the zero range potential is treated accurately by allowing it to vary freely with the kinetic energy part $\epsilon^{kin}$ of the energy per nucleon $\epsilon$ over the entire range of $\epsilon$. In a Fermi gas model of interacting neutrons and protons, the energy per nucleon for isospin asymmetric nuclear matter \cite{BCS08} is given by

\begin{equation}
 \epsilon(\rho,X) = [\frac{3\hbar^2k_F^2}{10m}] F(X) + (\frac{\rho J_v C}{2}) (1 - \beta\rho^n)  
\label{seqn1}
\end{equation}
\noindent
where $k_F$=$(1.5\pi^2\rho)^{\frac{1}{3}}$ which equals Fermi momentum in case of SNM, the kinetic energy per nucleon $\epsilon^{kin}$=$[\frac{3\hbar^2k_F^2}{10m}] F(X)$ with $F(X)$=$[\frac{(1+X)^{5/3} + (1-X)^{5/3}}{2}]$ and $J_v$=$J_{v00} + X^2 J_{v01}$, $J_{v00}$ and $J_{v01}$ represent the volume integrals of the isoscalar and the isovector parts of the M3Y interaction. The isoscalar $t_{00}^{M3Y}$ and the isovector $t_{01}^{M3Y}$ components of M3Y interaction potential are given by $t_{00}^{M3Y}(s, \epsilon)$=7999$\frac{\exp( - 4s)}{4s}$-$2134\frac{\exp( - 2.5s)}{2.5s}$+$J_{00}$(1-$\alpha\epsilon$)$\delta(s)$, $t_{01}^{M3Y}(s, \epsilon)$=-4886$\frac{\exp( - 4s)}{4s}$+$1176\frac{\exp( - 2.5s)}{2.5s}$+$J_{01}$(1-$\alpha\epsilon$)$\delta(s)$ $J_{00}$=-276 MeVfm$^3$, $J_{01}$=228 MeVfm$^3$, $\alpha=0.005$MeV$^{-1}$. The DDM3Y effective NN interaction is given by $v_{0i}(s,\rho, \epsilon) = t_{0i}^{M3Y}(s, \epsilon) g(\rho)$ where the density dependence $g(\rho) = C (1 - \beta \rho^n)$ and the constants $C$ and $\beta$ of the density dependence have been obtained from the saturation condition $\frac{\partial\epsilon}{\partial\rho} = 0$ at $X=0$, $\rho = \rho_{0}$ and $\epsilon = \epsilon_{0}$ where $\rho_{0}$ and $\epsilon_{0}$ are the saturation density and the saturation energy per nucleon, respectively, for the SNM \cite{BCS08}. The quantities $L$, $K_{sym}$ and $K_{asy}$ are defined and their values are calculated in Ref.\cite{CBS09}. It is worthwhile to mention here that the values listed in Table-1 of Refs.\cite{CBS09,Ch07} are for the approximate expression $K_{asy} \approx K_{sym}-6L$ for $K_\tau$. The third-order density derivative parameter $Q_0$ is given by \cite{Ch09} 

\begin{table*}[htbp]
\centering
\caption{Results of the present calculations (DDM3Y) of incompressibility of isospin symmetric nuclear matter $K_\infty$, nuclear symmetry energy at saturation density $E_{sym}(\rho_0)$, the slope $L$ and the curvature $K_{sym}$ parameters of the nuclear symmetry energy, the approximate isospin dependent part $K_{asy}$ and the exact part $K_\tau$ of the isobaric incompressibility (all in MeV) are compared with those obtained with RMF models \cite{Pi09}.}
\begin{tabular}{cccccccc}
\hline
\hline
Model&$K_\infty$&$E_{sym}(\rho_0)$&$L$&$K_{sym}$&$K_{asy}$&$Q_0$&$K_\tau$\\ 
\hline
 This work &$274.7\pm7.4$&$30.71\pm0.26$&$45.11\pm0.02$&$-183.7\pm3.6$&$-454.4\pm3.5$&$-276.5\pm10.5$&$-408.97\pm3.01$ \\ 
 FSUGold&230.0&32.59&60.5&-51.3&-414.3&-523.4&-276.77\\
 NL3&271.5&37.29&118.2&+100.9&-608.3&+204.2&-697.36 \\
 Hybrid&230.0&37.30&118.6&+110.9&-600.7&-71.5&-563.86\\ 
\hline
\hline
\end{tabular} 
\end{table*}
\noindent 

\begin{figure*}[htbp]
\vspace{5.0cm}
\eject\centerline{\epsfig{file=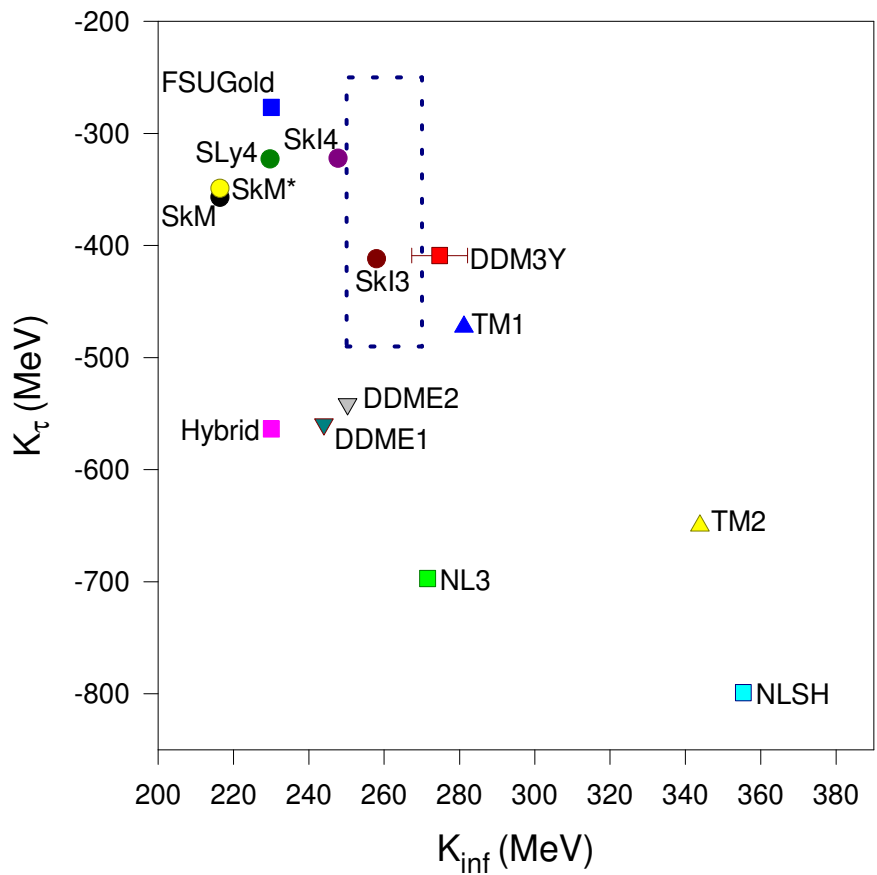,height=11.5cm,width=11.5cm}}
\caption
{The $K_\tau$ is plotted against  $K_\infty$ ($K_{inf}$) for the present calculation using DDM3Y interaction and compared with other predictions as tabulated in Refs.\cite{Pi09,Sa07}. The dotted rectangular region encompasses the values of $K_\infty=250-270$ MeV \cite{Sh09} and $K_\tau=-370\pm120$ MeV \cite{Ch09}.}
\label{fig1}
\vspace{0.0cm}
\end{figure*}

\begin{equation}
Q_0 = 27\rho_0^3 \frac{\partial^3 \epsilon(\rho,0)}{\partial {\rho^3}} \mid_{\rho=\rho_0}. 
\label{seqn2}
\end{equation}
\noindent
Using Eq.(1) one obtains

\begin{eqnarray}
&&\frac{\partial^3 \epsilon(\rho,X)}{\partial {\rho^3}} 
=-\frac{CJ_v(\epsilon^{kin})n(n+1)(n-1)\beta\rho^{n-2}}{2}  \nonumber \\ &&+\frac{8}{45}\frac{E^0_F}{\rho^3}F(X)(\frac{\rho}{\rho_0})^\frac{2}{3} 
+\frac{3\alpha JC}{5}n(n+1)\beta\rho^{n-1}\frac{E^0_F}{\rho} \nonumber \\ 
&&\times F(X)(\frac{\rho}{\rho_0})^\frac{2}{3} 
+\frac{\alpha JC}{5}[1-(n+1)\beta\rho^n] \frac{E^0_F}{\rho^2}F(X)(\frac{\rho}{\rho_0})^\frac{2}{3} \nonumber \\
&&-\frac{4\alpha JC}{45}[1-\beta\rho^n]\frac{E^0_F}{\rho^2}F(X)(\frac{\rho}{\rho_0})^\frac{2}{3}
\label{seqn3}
\end{eqnarray}
\noindent
where the Fermi energy $E^0_F$=$\frac{\hbar^2k_{F_0}^2}{2m}$ for the SNM at ground state,  $k_{F_0}$=$(1.5\pi^2\rho_0)^{\frac{1}{3}}$ and $J$=$J_{00}$+$X^2J_{01}$. Thus 

\begin{eqnarray}
&&\frac{\partial^3 \epsilon(\rho,0)}{\partial {\rho^3}} \mid_{\rho=\rho_0}
=-\frac{CJ_{v00}(\epsilon_0^{kin})n(n+1)(n-1)\beta\rho_0^{n-2}}{2} \nonumber \\  
&&+\frac{8}{45}\frac{E^0_F}{\rho_0^3}+\frac{3\alpha J_{00}C}{5}n(n+1)\beta\rho_0^{n-1} \frac{E^0_F}{\rho_0}+\frac{\alpha J_{00}C}{5} \nonumber \\   
&&\times [1-(n+1)\beta\rho_0^n]\frac{E^0_F}{\rho_0^2}-\frac{4\alpha J_{00}C}{45}[1-\beta\rho_0^n]\frac{E^0_F}{\rho_0^2}
\label{seqn4}
\end{eqnarray}
\noindent
where $\epsilon_0^{kin}$ is the kinetic energy part of the saturation energy per nucleon $\epsilon_0$. The calculations are performed using the values of the saturation density $\rho_0$=0.1533 fm$^{-3}$, the saturation energy per nucleon $\epsilon_0=-15.26\pm0.52$ MeV for the SNM and $n=\frac{2}{3}$ \cite{CBS09}. The saturation energy per nucleon is the volume energy coefficient $a_v$ of liquid drop model and the value of -15.26$\pm$0.52 MeV covers, more or less, the entire range of values obtained for $a_v$ for which the values of $C$ and $\beta$ are 2.2497$\pm$0.0420 and 1.5934$\pm$0.0085 fm$^2$ respectively \cite{BCS08}. Collisions involving $^{112}$Sn and $^{124}$Sn nuclei can be simulated with the improved quantum molecular dynamics transport model to reproduce isospin diffusion data from two different observables and the ratios of neutron and proton spectra. Constraints on the density dependence of the symmetry energy at subnormal density can be obtained \cite{Ts09} by comparing these data to calculations performed over a range of symmetry energies at saturation density and different representations of the density dependence of the symmetry energy. The results of the present calculations for $L$, $E_{sym}(\rho_0)$ and density dependence of $E_{sym}(\rho)$ \cite{CBS09} are consistent with these constraints \cite{Ts09}. In Table-1, the values of $L$, $E_{sym}(\rho_0)$, $K_{sym}$ and $K_\tau$ obtained using exact expression $K_\tau=K_{sym}-6L-\frac{Q_0}{K_\infty}L$ and its approximate form $K_{asy} \approx K_{sym}-6L$ are listed and compared with the corresponding quantities obtained with relativistic mean field (RMF) models \cite{Pi09}. 

      There seems to remain controversy over what is a reasonable value of incompressibility \cite{Sh06}. In the following we do not justify any particular value for $K_\infty$ but present our results in the backdrop of others for an objective view of the current scenario which, we stress, is still evolving. In Fig.1, $K_\tau$ is plotted against $K_\infty$ for the present calculation using DDM3Y interaction and compared with the predictions of FSUGold, NL3, Hybrid \cite{Pi09}, SkI3, SkI4, SLy4, SkM, SkM*, NLSH, TM1, TM2, DDME1 and DDME2 as given in Table-1 of Ref.\cite{Sa07}. The dotted rectangular region encompasses the recent values of $K_\infty=250-270$ MeV \cite{Sh09} and $K_\tau=-370\pm120$ MeV \cite{Ch09}. Although both DDM3Y and SkI3 are within the above region, unlike DDM3Y the $L$ value for SkI3 is 100.49 MeV which is much above the acceptable limit of 45-75 MeV \cite{Wa09} whereas DDME2 which gives $L=51$ MeV is reasonably close to the rectangular region. It is worthwhile to mention here that the DDM3Y interaction with the same ranges, strengths and density dependence which gives $L=45.11\pm0.02$ here, provides good descriptions of scattering (elastic and inelastic), proton radioactivity \cite{BCS08} and $\alpha$ radioactivity of superheavy elements \cite{CSB06,scb07}. The present NSE is `super-soft' because it increases initially with nucleonic density up to about two times the normal nuclear density and then decreases monotonically (hence `soft') and becomes negative (hence `super-soft') at higher densities (about 4.7 times the normal nuclear density) \cite{BCS08,CBS09} and is consistent with the recent evidence for a soft NSE at suprasaturation densities \cite{Zh09} and with the fact that the super-soft nuclear symmetry energy preferred by the FOPI/GSI experimental data on the $\pi^+/\pi^-$ ratio in relativistic heavy-ion reactions can readily keep neutron stars stable if the non-Newtonian gravity proposed in the grand unification theories is considered \cite{We09}. 

      In summary, we conclude that the approximate expression $K_{asy} \approx K_{sym}-6L$ which is quite often used in place of $K_\tau=K_{asy}-\frac{Q_0}{K_\infty}L$ can lead to a difference of about ten percent (DDM3Y) or more (FSUGold) in $K_\tau$. The recently measured data on the breathing mode of Sn isotopes seem to favour a constraint $K_\tau= -550\pm100$ MeV for the asymmetry term in the nuclear incompressibility \cite{Li07,Ga07}. First and foremost, $K_\tau$ should not be inferred from an extrapolation to the $A \rightarrow \infty$ limit from laboratory experiments on finite nuclei. Rather, one should continue to follow the procedure advocated by Blaizot \cite{Bl80,Bl95} and demand that the values of both $K_\infty$ and $K_\tau$ be those predicted by a consistent theoretical model that successfully reproduces the experimental giant monopole resonance (GMR) energies of a variety of nuclei. We reiterate that in the present contribution, both $K_\infty$ and $K_\tau$ refer to the bulk properties of the infinite system. Nevertheless, considering the fact that the extracted value of $K_\tau=-550\pm100$ MeV \cite{Li07} is from GMR of nuclei as light as Sn isotopes, the present value $-408.97\pm3.01$ MeV is in reasonably close agreement whereas it is in excellent agreement with $K_\tau=-389\pm12$ MeV (NL3),$-345\pm12$ MeV (SVI2),$-395\pm13$ MeV (SIGO-c) \cite{Sh09} when extracted reproducing GMR energies of nuclei such as $^{208}$Pb, Sn isotopes and $^{90}$Zr among others.

\end{document}